# Selection rules for Brillouin light scattering from eigenvibrations of a sphere

**Y. Li, H. S. Lim, S. C. Ng, Z. K. Wang, M. H. Kuok[*]**

*Department of Physics, National University of Singapore*

*2 Science Drive 3, Singapore 117542, Singapore*

**Abstract**

Selection rules governing Brillouin light scattering from the vibrational eigenmodes of a homogeneous, free-surface submicron sphere have been derived using group theory. The derivation is for the condition where the sphere diameter is of the order of the excitation light wavelength. Well-resolved spectral data obtained from Brillouin light scattering from submicron silica spheres provide experimental verification of the selection rules.

[*]Corresponding author. Tel: +65 65162604; Fax: +65 67776126; E-mail address: phykmh@nus.edu.sg (M. H. Kuok)

The acoustic eigenmodes of a homogeneous sphere with a free surface have been studied by Lamb [1] who classified them into two categories, *viz*. spheroidal and torsional modes, labeled by the angular momentum quantum number $l$, where $l = 0, 1, 2, \ldots$ for spheroidal modes, and $l = 1, 2, 3, \ldots$ for torsional modes. The sequence of eigenmodes, in increasing order of energy, is indexed by $n$ (= 1, 2, 3, . . .). Lamb's theory predicted that the frequencies of the eigenmodes $(n, l)$ scale as $1/d$, where $d$ is the sphere diameter. Hence, vibrations of nanometer-size spherical particles are usually investigated by Raman scattering [2-6] while those of micron-size ones by Brillouin light scattering [7-12]. However, it is to be noted that the distinction between Raman scattering and Brillouin light scattering is artificial, as the inelastic light scattering mechanism from eigenvibrations of a sphere is the same. Only the detection technique, which depends mainly on the size of the sphere concerned, is different.

Much controversy surrounds the inelastic light scattering selection rules for the vibrational eigenmodes of a sphere. Duval derived the Raman selection rules for the condition in which the sphere diameter $d \ll \lambda$, the wavelength of the excitation light (usually ~ 500 nm) [13]. He predicted that only spheroidal modes with $l = 0, 2$ are Raman-active while torsional modes are not observable by Raman scattering. The same results were obtained by Montagna and Dusi, and Li *et al.* from calculations of the Raman coupling coefficients of the acoustic vibrations [14,15]. However the recently derived Raman selection rules of Kanehisa preclude the observation of spheroidal modes and state that only the torsional mode with $l = 2$ is Raman-active [16]. The Comment, by Goupalov *et al.* [17], which refuted his model and his subsequent rebuttal [18] have exacerbated the controversy. Interestingly, Tanaka *et al.* claimed that both spheroidal modes of even $l$, and torsional modes of odd $l$ are Raman-active [6].

No selection rules governing Brillouin light scattering $(d \sim \lambda)$ from confined acoustic modes in a sphere have, to date, been reported. Consequently, in previous studies, mode assignments of Brillouin spectra of submicron or micron-size silica, polystyrene and $CaCO_3$ spheres were not based on any one consistent set of selection rules [7-12]. In some cases, observed spectral peaks were assigned as spheroidal modes with $l = 0, 2, 4, \ldots$, [7-10] while in others, as spheroidal modes with $l$ = any arbitrary integer [11,12]. To address this problem, we present a derivation, from group theory, of the Brillouin selection rules and their experimental verification.

The selection rules for inelastic light scattering from confined acoustic modes in a homogeneous sphere are now discussed. The elastic displacements $\boldsymbol{u}$ of the modes are given by

$$\partial_t^2 \boldsymbol{u} - V_L^2 \nabla(\nabla \cdot \boldsymbol{u}) + V_T^2 \nabla \times \nabla \times \boldsymbol{u} = 0, \tag{1}$$

where $V_L$ and $V_T$ are the bulk longitudinal and transverse velocities respectively. The elastic displacements can be categorized into torsional and spheroidal displacements which are given respectively by

$$\boldsymbol{u}_t = \nabla \times (\boldsymbol{r}\psi) \tag{2}$$

$$\boldsymbol{u}_s = \nabla \chi + \nabla \times \nabla \times (\boldsymbol{r}\phi), \tag{3}$$

where $\psi$, $\chi$ and $\phi$ are the scalar potentials which satisfy the scalar Helmholtz equation, and are expressed in terms of Bessel functions $j_l$ and spherical harmonics $Y_{lm}$ in spherical coordinates. The symmetry group of the sphere is the group, of the proper and improper rotations, that is isomorphic to the full rotation group $\mathbf{O}(3)$. Their irreducible representations are $D_g^{(l)}$ and $D_u^{(l)}$, where $l = 0, 1, 2, 3,\ldots$, and $g$ and $u$ denote even parity and odd parity respectively [19]. The spheroidal vibrations transform as the irreducible representations $D_g^{(0)}$, $D_u^{(1)}$, $D_g^{(2)}$,… , while the torsional vibrations as $D_g^{(1)}$, $D_u^{(2)}$, $D_g^{(3)}$,… of $\mathbf{O}(3)$ [13].

In Raman scattering, the sphere diameter is much shorter than the excitation light wavelength. Hence, the phase of the excitation light within the sphere can be assumed to be constant, and only the electric dipole moment need be considered in the light scattering process. The irreducible representations of the components of the dipole moment $\sum_i e_i r_i$ transform as $D_u^{(1)}$. The operator involved in inelastic light scattering is the symmetric polarizability tensor $\alpha_{ij}$ ($\alpha_{ij} = \alpha_{ji}$) whose components transform as the irreducible representations resulting from the symmetric product [13]:

$$[D_u^{(1)} \times D_u^{(1)}]_{\text{symmetric}} = D_g^{(0)} + D_g^{(2)}. \tag{4}$$

Duval thus concluded from Eq. (4) that only spheroidal modes with $l = 0, 2$ are observable by Raman spectroscopy.

However, if the diameter of the sphere is comparable to the incident light wavelength ($d \sim \lambda$), the frequencies of the vibrational modes of sphere lie in the gigahertz range, and its modes are thus detectable by Brillouin light scattering. In this case, the assumption that the phase of the incident light within the sphere is constant is no longer valid. In Mie scattering, where the sizes of the particles can be larger than the excitation light wavelength, besides the electric dipole moment, higher-order terms in the electric multipole expansion are also taken into account in the series solution of the scattering theory [20]. Similarly, besides the electric dipole moment, these higher-order terms are also taken into consideration in our derivation of Brillouin selection rules.

The components of the electric multipole tensors, of rank $k$, transform as the irreducible representations $D^{(k)}$ of the full rotation group **O**(3) [19]. Thus, for example, the rank 2 multipole tensor, which is the electric quadrupole tensor, has the irreducible representation $D^{(2)}$. As before, the operator involved in the inelastic light scattering is the symmetric polarizability tensor whose components transform as the irreducible representations that are obtained from the symmetric product [21]

$$[D^{(k)} \times D^{(k)}]_{\text{symmetric}} = D_g^{(0)} + D_g^{(2)} + D_g^{(4)} + ... + D_g^{(l)}, \tag{5}$$

where $k = 1, 2, 3, ...,$ and $l = 0, 2, 4, ..., 2k$. Therefore, it follows that for $d \sim \lambda$, only the spheroidal modes of a sphere are observable by Brillouin spectroscopy. Unlike Duval's selection rules for Raman scattering, spheroidal vibrations with $l = 0, 2, 4, ...$ are Brillouin active. Torsional modes are not detectable by inelastic light scattering.

Brillouin light scattering from six samples of loose matrix-free monodisperse amorphous silica spheres, with diameters ranging from 140 – 800 nm, had been measured in the 180°-backscattering geometry using a 6-pass tandem Fabry-Perot interferometer [8]. The 514.5 nm line of an argon-ion laser was used to excite the spectra. The measured spectra contain well-resolved sharp peaks as can be seen in the representative spectrum of the 360nm-diameter sphere sample displayed in Fig. 1. The variations of the measured frequencies of the eigenmodes of the various spheres with inverse sphere diameter are presented in Fig. 2.

These experimental data were used to ascertain the validity of the selection rules derived above based on the following approach. In this approach, the measured frequencies of the two lowest-energy modes, of the six sphere sizes studied, were used in the evaluation of the velocities $V_L$ and $V_T$. First, a trial assignment of the observed lowest and second lowest energy modes was made. Second, the parameters $V_L$ and $V_T$ were determined by least-squares fitting the data to the Lamb theory [1]. This was done by minimizing the residual $\sum_{i=1}^{N}(\nu_i^{th} - \nu_i^{expt})^2$ by varying $V_L$ and $V_T$ within the reasonable ranges of $2000 \leq V_L \leq 7000$ m/s and $1000 \leq V_T \leq 5000$ m/s. Here $\nu^{th}$ and $\nu^{expt}$ are the theoretical and measured frequencies respectively, and $N$ (= 12) the total number of mode frequencies of the two lowest-energy vibrations for the six sphere sizes studied. Third, using the fitted values of $V_L$ and $V_T$, the frequencies of the two lowest-energy modes were calculated to check for self-consistency i.e. agreement between the resulting energy ordering and the trial assignments. Finally, the frequencies of the higher-energy modes are computed for the assignments that meet this self-consistency condition.

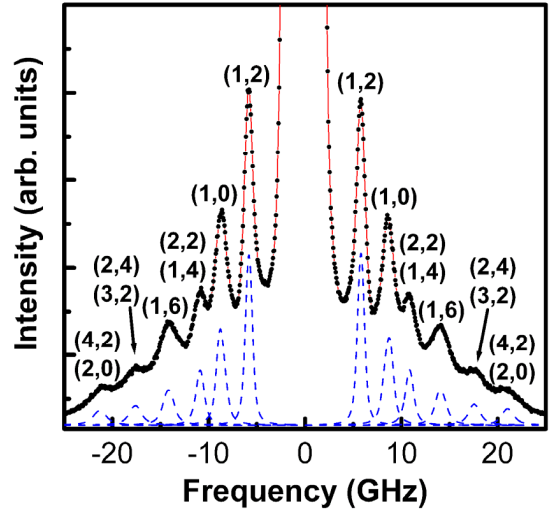

**Fig. 1**. Brillouin spectrum of 360nm-diameter silica spheres. The experimental data are denoted by dots and Brillouin peaks were fitted with Lorentzian functions shown as dashed curves. The assignment of the confined acoustic modes, labeled by ($n$, $l$), is based on our selection rules as described in the text.

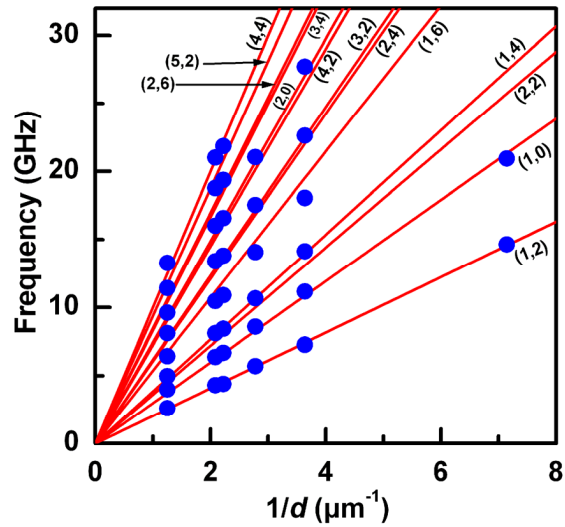

**Fig. 2**. Dependence of frequencies of confined acoustic modes ($n$, $l$) in silica microspheres on inverse sphere diameters ($d$). Experimental data are denoted by dots. The measurement errors are the size of the dots shown. Solid lines represent theoretical frequencies of spheroidal modes with $l = 0, 2, 4,…$, calculated based on Lamb's theory and our selection rules.

According to our selection rules, only even-$l$ spheroidal modes are observable by inelastic light scattering. Hence, when making the trial assignment, all possible pairs of $(n, l)$ for $n = 1, 2$ and $l = 0, 2, 4$ were considered. This is because our calculations reveal that the two lowest-energy modes do not have $n$ and $l$ higher than these values for $V_L$ and $V_T$ within the same ranges specified above. It was found that almost all the possible choices of $(n, l)$ do not satisfy the self-consistency condition. For instance, the trial assignment of (1,0) and (1,4) yields, in order of increasing energy, the sequence (1,2), (1,0), (2,2), (1,4), … in contradiction to the trial assignment.

It turns out that only two trial assignments meet this self-consistency requirement. One of them is (1,0) and (1,2) as the respective lowest and second lowest-energy modes. Using the fitted $V_L$ and $V_T$ based on this assignment, the frequencies of the higher-energy modes with even $l$ were calculated. However this assignment is unacceptable as there is discrepancy between the calculated and measured frequencies of the higher-energy modes. For the other trial assignment of (1,2) and (1,0), the frequencies of the corresponding higher-energy modes with even $l$ were computed in a similar fashion. Results of the calculations are presented in Fig. 2. The (4,4) spheroidal vibration has the highest energy. It should be emphasized that *all* modes, both measured and theoretical, with lower energies are also displayed. Figure 2 clearly illustrates the good agreement between theory and experiment. Additionally, a one-to-one correspondence exists between them i.e. not only are all observed modes accounted for, but also there is no omission of any theoretical modes nor are there any theoretical modes unaccounted for. Hence, our Brillouin selection rules are experimentally confirmed.

The fitted values of $V_L$ and $V_T$ are 3885 and 2436 m/s respectively. These are reasonable values, as they are expected to be lower compared to $V_L \approx 5500$ and $V_T \approx 3000$ m/s for bulk silica, due to low dimensionality and the possible presence of defects as discussed in Ref. 8. Also, these fitted velocities obtained from the confined eigenmodes are consistent with the respective $V_L$ and $V_T$ values of 3778 and 2518 m/s measured by Lim *et al.* in their study of Brillouin light scattering from bulk acoustic waves in 1.45 – 3.95 µm silica spheres [8].

In summary, the Brillouin selection rules for a sphere with a diameter of the order of the excitation light wavelength have been derived from group theory. In the derivation, besides the electric dipole moment, higher-order terms in the electric multipole expansion have also been taken into account. It is found that only the spheroidal modes with $l$ = 0, 2, 4, … are observable by Brillouin spectroscopy. Torsional modes are not detectable by inelastic light scattering. Our selection rules, for $d \sim \lambda$, allow the observation of confined acoustic modes with even values of $l \geq 4$, in contrast to Duval's Raman rules, for $d \ll \lambda$, for which $l$ = 0, 2 only. Our selection rules were experimentally confirmed by spectral data on inelastic (Brillouin) light scattering from silica microspheres.

**Acknowledgement**

Funding from the Ministry of Education, Singapore under research grant R-144-000-185-112 is gratefully acknowledged.